\documentclass[referee,sn-basic]{sn-jnl}

\usepackage{graphicx}%
\usepackage{multirow}%
\usepackage{amsmath,amssymb,amsfonts}%
\usepackage{amsthm}%
\usepackage{mathrsfs}%
\usepackage[title]{appendix}%
\usepackage{xcolor}%
\usepackage{textcomp}%
\usepackage{manyfoot}%
\usepackage{booktabs}%
\usepackage{algorithm}%
\usepackage{algorithmicx}%
\usepackage{algpseudocode}%
\usepackage{listings}%

\theoremstyle{thmstyleone}%
\theoremstyle{thmstyletwo}%

\theoremstyle{thmstylethree}%

\begin{document}

\title[Wind Pattern at Merak]{Study of Wind pattern at the incursion site of Pangong Tso near Merak Village}


\author*[1]{\fnm{Belur} \sur{Ravindra}}\email{ravindra@iiap.res.in}

\author[1]{\fnm{Deepangkar} \sur{Sarkar}}\email{sdeepangkar@gmail.com}
\equalcont{These authors contributed equally to this work.}

\author[1]{\fnm{Shantikumar Singh} \sur{Ningombam}}\email{shanti@iiap.res.in}
\equalcont{These authors contributed equally to this work.}

\author[1]{\fnm{Stanzin} \sur{Tundup}}\email{stanzin@iiap.res.in}
\equalcont{These authors contributed equally to this work.}

\author[1]{\fnm{Namgyal} \sur{Dorje}}\email{namgyal@iiap.res.in}
\equalcont{These authors contributed equally to this work.}

\author[1]{\fnm{Angchuk} \sur{Dorje}}\email{angchuk@iiap.res.in}
\equalcont{These authors contributed equally to this work.}

\author[3]{\fnm{Prabhu} \sur{Kesavan}}\email{kes.prabhu@gmail.com}
\equalcont{These authors contributed equally to this work.}

\author[1,2]{\fnm{Dipankar} \sur{Banerjee}}\email{dipu@aries.res.in}
\equalcont{These authors contributed equally to this work.}

\affil*[1]{\orgname{Indian Institute of Astrophysics}, \orgaddress{\street{II Block, Koaramangala}, \city{Bengaluru}, \postcode{560034}, \state{Karnataka}, \country{India}}}

\affil[2]{\orgname{Aryabhatta Research Institute of Observational Sciences}, \orgaddress{\street{Manora Peak}, \city{Nainital}, \postcode{263001}, \state{Uttarakhand}, \country{India}}}

\affil[3]{\orgaddress{\street{Nanjanadu}, \city{The Nilgiris}, \postcode{643004}, \state{Tamil Nadu}, \country{India}}}

\abstract{This study analyzes twelve years of wind speed and direction data collected at the proposed National Large Solar Telescope (NLST) site near Pangong Tso, Merak village, Leh-Ladakh. A weather station from Campbell Scientific Instruments, installed in 2008, has been continuously monitoring meteorological parameters, including wind speed and direction. The data reveals a consistent pattern of predominantly northwest winds, particularly during morning hours, with speeds generally below 5 m/s. While seasonal variations influence wind speed and direction, the overall trend remains stable. To assess the site's suitability for astronomical observations, we compared high-altitude wind speeds at various renowned astronomical sites using reanalysis data from 2008 to 2020. Strong correlations were observed between surface and high-altitude wind speeds at 10~m, 50~m, and 500~m. Statistical analysis of 200-mbar pressure level wind speeds identified La Palma as the most favorable site with a wind speed of 18.76~m/s. La Silla, on the other hand, exhibited the highest wind speed at 34.76~m/s. Merak's estimated wind speed of 30.99~m/s, coupled with its favorable wind direction and low surface wind speeds, suggests its potential as a promising site for astronomical observations.}

\maketitle

\section{Introduction}\label{sec1}

The present period of stellar astronomy has seen telescopes larger than 10-m class \citep{1995PASP..107..994D}. Solar astronomy is also in competition to achieve high-resolution imaging of the sun \citep{2020SoPh..295..172R}. The prevailing meteorological conditions at the observing site uniquely decide the image quality of any telescope. The meteorological parameters, such as wind speed, direction, humidity, etc., will affect the quality of the solar imaging made by the telescopes. These meteorological parameters also play a significant role in the construction of the building, dome, and even the telescope's structure to some extent. The operation of the telescope, dome, and adaptive optics also depends on the local meteorological conditions.

Generally, the high-speed wind degrades the image quality by shaking the telescope and putting a load on the telescope mount and the mirrors. The random load on the telescope mount and other nearby structures will cause tracking and pointing errors, resulting in blurred images. The high-speed winds generate turbulence, which degrades the image quality \citep{app13106354}. Besides, it transports the aerosol content to long distances, which will augment the risk of deposition of tiny particles on the mirror and other reflecting surfaces of the telescope.

The air turbulence occurs when air at different temperatures and densities mixes, causing fluctuations in air density. This can lead to rapid image motion, image distortion, and intensity fluctuations, all of which can degrade image quality. Historically, astronomers have often assessed seeing conditions by measuring the width of a star's image. However, this measurement can be influenced by factors other than atmospheric turbulence, such as telescope imperfections or vibrations. A more precise measure of atmospheric turbulence is Fried's parameter, r$_{0}$, which represents the diameter of the turbulent eddy that significantly affects wavefront distortion \citep{1966JOSA...56.1372F}. It is calculated based on the refractive index structure constant, C$_{n}^{2}$, which quantifies the strength of turbulence at different altitudes. The Kolmogorov model of turbulence describes how turbulence affects physical quantities like temperature and air density.  A larger r$_{0}$ indicates less atmospheric turbulence, leading to better seeing conditions. Essentially, r$_{0}$ represents the maximum telescope aperture size that can achieve diffraction-limited resolution, unaffected by atmospheric distortion.

The quality of the image of a telescope is decided by the parameter called `seeing' at the site. The value of Fried's parameter depends on the local topographic conditions, weather patterns, aerosol contents, and other developmental activities at the site. Hence, measuring this parameter for the site where the telescope is intended to be put is essential. Measuring this parameter for a prolonged time will help design the adaptive optics system, which is now an integral part of any large telescopes making high-angular resolution observations. It also helps to understand how the local weather pattern affects the value of seeing and how the seasonal variations in the local condition affect the high-resolution observations.

A study by \cite{2006PASP..118..172B} revealed that the distribution of r$_{0}$ is influenced by wind direction. Specifically, they observed a degradation in seeing conditions when the wind blows between 60$^{\circ}$ and 90$^{\circ}$ North. Additionally, they found that increased wind speed correlates with poorer seeing quality. This is attributed to the enhancement of atmospheric turbulence caused by increased wind shear at higher wind speeds. \cite{2003A&A...400.1163T} explored the relationship between wind speed and C$_N^{2}$ with altitude. Furthermore, \cite{2006SPIE.6267E..0LD} reported that r$_{0}$ values typically range between 5 and 8~cm when wind speeds are between 4 and 7~m/s. Deviations from this optimal range, either above 7~m/s or below 4~m/s, lead to a deterioration in r$_{0}$.

The Changthang region of eastern Ladakh has uniqueness because of the high-altitude environment and the large number of clear skies. Indian Institute of Astrophysics (IIA) has carried out visual inspection of cloud data, meteorological parameters and other astronomical parameters since the early 1990s, as part of an astronomical site survey program to set up astronomical facilities in the optical and NIR regions {\citep{1989BASI...17...83S, 2000BASI...28..441B}}. After investigating several years of cloud data, meteorological parameters, and other astronomical parameters, IIA has installed a 2-m Himalayan Chandra Telescope (HCT), operating in the optical and near-infrared region, at Indian Astronomical Observatory (IAO) Hanle ($32^\circ46^\prime46^{\prime\prime}$N;$78^\circ57^\prime52^{\prime\prime}$E; 4500 m, amsl) in 2000 \citep{2014PINSA..80..887P}. The number of clear skies in a year at IAO-Hanle and Merak is about 68-78\% and 61-68\%, respectively, as reported by \cite{2021MNRAS.507.3745N}.

As an additional astronomical facility, IIA has also planned to install a 2-m telescope dedicated to the observations of the sun at a very high resolution. The telescope will be the Gregorian type with symmetric optics on the alt-azim mount. With the minimum reflecting surfaces, the telescope provides a high throughput at the final focus \citep{2012ASPC..463..395H}. The open dome design is intended to provide a free flow for the wind. The structure of the building and the shape of the dome will be based on the wind direction and speed. Among the Merak and Hanle sites, Merak is more suitable for the proposed solar telescope as the site is located at the incursion site of Pangong Lake. Several instruments were installed at Merak in early 2008 as a part of this project. One of the instruments is the automatic weather station installed at the site in 2008. The instrument takes the wind speed and direction data, air temperature, and relative humidity measurements. Continuous data is available Except for about a year of data gaps. More than ten years of data is sufficient to study the variations in the wind pattern that occurred at the site due to natural and anthropogenic causes.

This study is based on the wind data collected by the automatic weather station instrument installed at the Pangong Tso (lake) incursion site. We also studied upper air wind data from reanalysis data to strengthen the study. We used twelve years of collected data to look for the wind pattern at the site. This study will provide information on whether there is any change in the wind pattern over the year, how the wind pattern changes over the season, etc. The paper is organized as follows. In the next section (Section~\ref{sec:2}), we give brief details about the topographical structure near the site. In Section~\ref{sec:3}, we provide information on the instrument, data structure, and analysis method. In Section~\ref{sec:4}, the results of the study are presented. The paper ends with a summary of the results and discussions.

\section{The site}
\label{sec:2}

\begin{figure}[!h]
\centering
\includegraphics[width=0.7\textwidth]{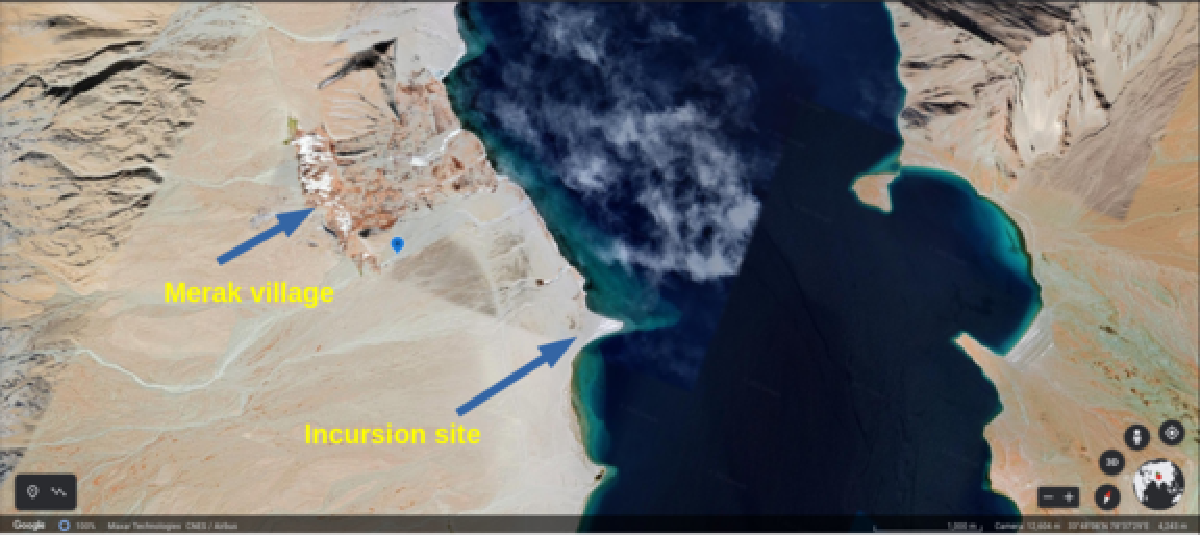}
\caption{A Google Earth image of the incursion site showing the Pangong Tso lake, Merak village and hills. The North-East direction and the scale are also shown in the map.}
\label{fig:1}
\end{figure}

The entire region in Ladakh is located at a high elevated ($>$3000 m masl) mountainous region. Further, the whole region is not affected by the Indian monsoon as the region is located in the rain shadow area of the Himalayas. The incursion site of the proposed solar telescope project is located between the Zanskar range in the south, the Karakoram range in the north, and the Tibetan Plateau in the east. The nearby village of the site is called Merak, with a population of 300 people. It is located about 118 km from IAO-Hanle and about 103 km from Leh town (aerial distance), the capital of Ladakh Union Territory.

The width of the Pangong lake varies from 3 to 7 km and is about 134 km long. Due to the large water bodies, the incursion site behaves like a small island above an elevated high-rise mountain range. The site is located far from any big town/city, which is attributed to low aerosol ($\sim$0.05 at 500 nm) content at the site {\citep{2015Env....22..21}.} During the summer, the daytime temperature is about 20$^{\circ}$C; in the winter, it is below --10$^{\circ}$C. The lake is running in the North-West and South-East direction. The water from the glacier in the mountains feeds the lake during the summer. The top layer of the lake is frozen during the winter. The site is cold and dry in all seasons and falls in the range of the Changtang Cold desert. Figure~\ref{fig:1}(a) shows the location of the incursion in the Google map.

\section{Data collection and analysis}
\label{sec:3}

\begin{figure}[!h]
\centering
\includegraphics[width=0.7\textwidth]{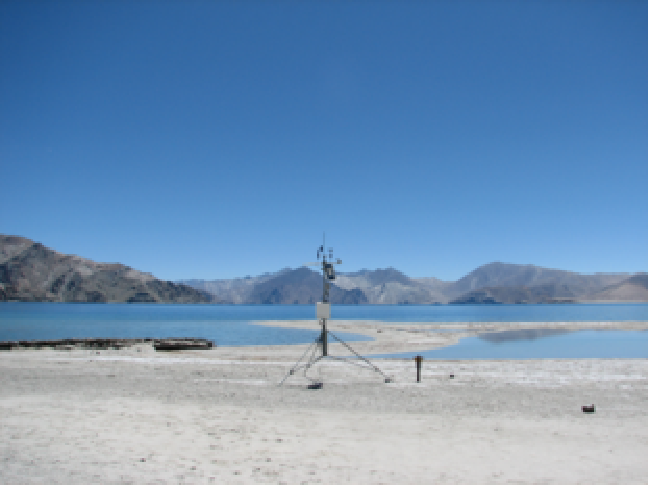}
\caption{A picture of automated weather station installed at the incursion site.}
\label{fig:2}
\end{figure}

The instrument used to collect the meteorological parameters is the automated weather station (AWS). This integrated device consists of 8 instruments supplied by the Cambell Scientific Instruments. The sensors measure the air and soil temperature and humidity are placed on a 3-m height pole. A three-cup anemometer and a wind vane mounted on a small cross arm on the bar. The rain gauge, barometric pressure sensor, and pyranometer to quantify solar radiation are also present. All these instruments are assembled on a single rod mounted on a tripod. Each device has sensors, and the power is supplied to each sensor through a rechargeable battery mounted in a closed box. The battery is recharged through the solar panel. Figure~\ref{fig:2} shows the picture of the AWS installed at the incursion site. The instrument sensors are located at a height of 3-m above the ground. The rain gauge and soil temperature sensors are kept on the ground.

The three-cup anemometer is utilized in the measurement of wind speed. The 40~mm diameter hemispherical cup is placed on a wheel assembly of 12~cm diameter. The rotation of the cup produces the sine wave signal with a frequency corresponding to the wind speed. The vane senses the wind direction, and the voltage produced by the sensor is proportional to the wind direction. The anemometer can measure the wind speed ranging from 0 to 50~m~s$^{-1}$.  The instruments operate optimally within an ambient temperature range of --20$^{\circ}$C to $+$60$^{\circ}$C. They offer a high level of accuracy, measuring wind speed with a precision of $\pm$0.11~m/s and a threshold speed of 0.45~m/s. Wind direction is measured with an accuracy of $\pm$5$^{\circ}$. These measurement accuracies are comparable to those reported in \cite{2015PASP..127.1292Z, 2006PASP..118..172B}, which range from $\pm$0.2 to $\pm$3~m/s for wind speed and $\pm$2$^{\circ}$ to $\pm$3$^{\circ}$ for wind direction. The sensors are connected to the data logger (CR-1000 model), which communicates between the instruments, converts the electrical signal into respective measurement values, and the data is stored in the device's memory. The data can be transferred to the computer whenever required. The data is stored at every 10-second interval and in tabular form, including the observations' date and time.

The AWS was installed at the incursion site in June 2008. It is continuously recording the data since then. However, the instrument was moved to Hanle in April 2015 and installed back at Merak in November 2016. Hence, there is no data available during this period. We have not analyzed the 2016 data because only less than two months of data is available. But, we have used the 2015 data in the analysis. The data is continuously taken till today. The data used for the analysis is obtained from 2009 to 2021. In this paper, we show the daily data plots, monthly averaged, and yearly averaged data plots. Further, to study the upper air wind profile and turbulence profile, we used global atmospheric reanalysis data of the Modern-Era Retrospective analysis for Research and Applications, version 2.0 \citep[MERRA-2;][]{Gelaro2017}. The data produced is the latest atmospheric reanalysis of the modern satellite era produced by NASA’s Global Modeling and Assimilation Office (GMAO) and has a spatial resolution of 0.5$^{\circ}$ by 0.625$^{\circ}$ at 73 vertical pressure levels. Further, we used ERA-5 reanalysis data which is the fifth generation reanalysis of the European Centre for Medium-Range Weather Forecasts (ECMWF) with a spatial resolution of 0.25$^{\circ}$ by 0.25$^{\circ}$. The details of the data products are described by \cite{Hersbach2020}.

\section{Results}
\label{sec:4}
\subsection{Windrose Diagrams}
The wind rose plots are used to show the distribution of the direction of the wind flow and the speed at a particular location. It has a circular pattern (polar plots) plotted at regular intervals representing the frequency of time of the wind blowing from specific directions. Each circle represents a different frequency, which starts from zero at the center and increases in frequency as it goes to the bigger circles. The circular pattern is divided into four quadrants, each representing the direction of the North, South, East, and West. This can be further divided into North-East (NE), North-West (NW), South-East (SE) and South-West (SW). Each can be again divided into NNE, ENE, ESE, SSE, and so on, 16 cardinal directions. The petals or spoke-like patterns in a circle show the speed and direction of the wind, which indicates how much of the time the wind blows from a particular direction. Several petals indicate the percentage of the time the wind blows from that direction with a specific speed range. The color code represents the speed range.

\begin{figure*}[h!]
\centering
\includegraphics[width=0.9\linewidth]{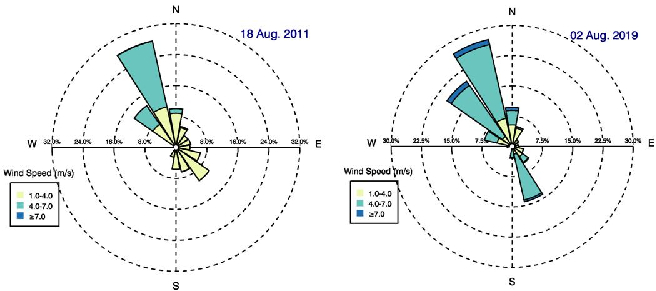}
  \caption{Windrose plot showing the wind speed and direction for a single day during August 18, 2011 (left) and  August 02, 2019(right). A large fraction of the time the wind speed was below 7~m~s$^{-1}$.}
  \label{fig:3}
\end{figure*}

A typical one-day plot of wind speed and wind direction in the form of windrose is shown in Figure~\ref{fig:3}. The left-side plot is for August 18, 2011, and the right-side plot is for August 02, 2019. While making the windrose plots, we considered only points with a speed larger than 1~ms$^{-1}$. For the plots, the wind data taken between 6:00 AM and 6:00 PM are used. In both plots, the general direction of the wind flow is in NW most of the time. But, for a short time, the wind also blows from South-East direction. On August 18, 2011, about 52\% of the time, the wind blew in the NW direction, and about 20\% of the time, it was in the SE direction. About 28\% of the time, the wind speed was between 4 -- 7~ms$^{-1}$. On August 02, 2019, about 55\% and 18\% of the time, the wind blew from NW and SE direction, respectively. The wind speed was between 4 -- 7~ms$^{-1}$ most of the time.

\begin{figure*}[h!]
\centering
  \includegraphics[width=\linewidth]{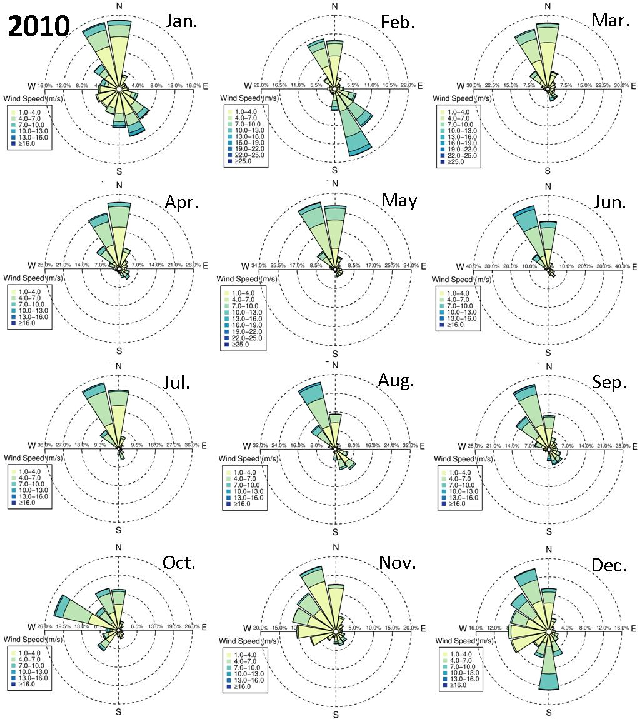}
  \caption{Seasonal variation in the wind speed and direction at Merak for the year 2010.}
  \label{fig:4}
\end{figure*}

\begin{figure*}[h!]
\centering
  \includegraphics[width=\linewidth]{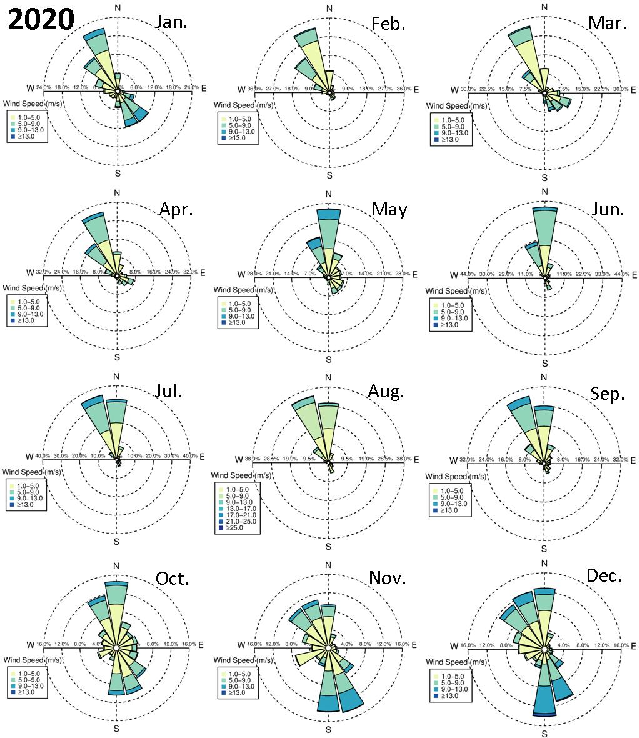}
  \caption{Seasonal variation in the wind speed and direction at Merak for the year 2020.}
  \label{fig:5}
\end{figure*}

Figures~\ref{fig:4} and \ref{fig:5} show the direction of the wind pattern and speed in each month of 2010 and 2020, respectively. The length of each petal in the circle indicates how often the wind blew from that direction. The wind pattern is computed by averaging the speed and direction over a month but keeping the daytime only from 6:00 AM to 6:00 PM. Generally, the wind blows from the N-W direction. Except for December, January, and February, more than 50\% of the time, the wind blew from the NW direction with a speed of less than 7~ms$^{-1}$ at least for 50\% of the time. In the rest of the months, apart from NW, the wind also blew from SE, and about 5-10\% of the time, the speed was also larger than 7~ms$^{-1}$. Almost similar results were observed in the year 2020. The color code on the lower left side of each wind-rose plot provides the speed in m~s$^{-1}$. In order to find how many hours over which the particular wind speed is observed, it just needs to multiply the corresponding frequency by the exact amount of time. In November, December, and January, the wind blows not only in the N-W and S-E but a small fraction of the time, it blows in other directions as well, S-W and N-E. In order to see whether the same pattern is followed in all the years, we have plotted the annual average wind speed.

\begin{figure*}[h!]
  \includegraphics[width=\linewidth]{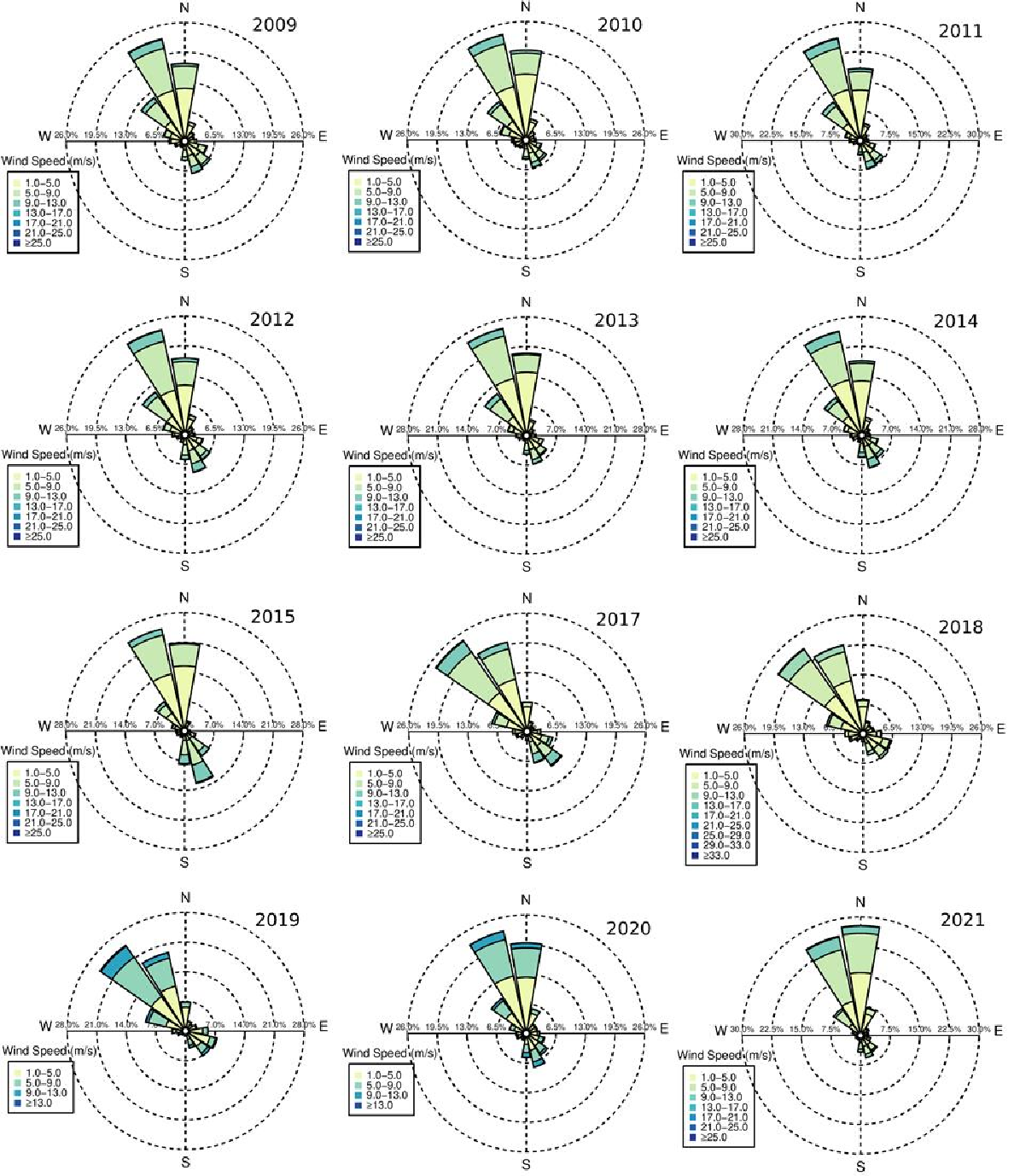}
  \caption{Annual variation of wind direction and speed from the year 2009 to 2021.}
  \label{fig:6}
\end{figure*}

Figure~\ref{fig:6} shows the annual distributions of wind speed and direction from 2009 to 2021. These were plotted for all months' data for a time period of 6~AM to 6~PM each day to study the annual behavior of the wind pattern. The yearly distribution of the wind speed and direction plot shows more than 60\% of the time in North and North-Westerly and less than 20\% in South and South-Easterly. Apart from this, a small fraction of the time, the wind blows in other directions. In all the years, about 40 -- 50\% of the time, the wind speed was less than 5~ms$^{-1}$, and about 25\% of the time, the wind speed was less than 9~ms$^{-1}$.

\begin{figure}[h!]
  \includegraphics[width=1\linewidth]{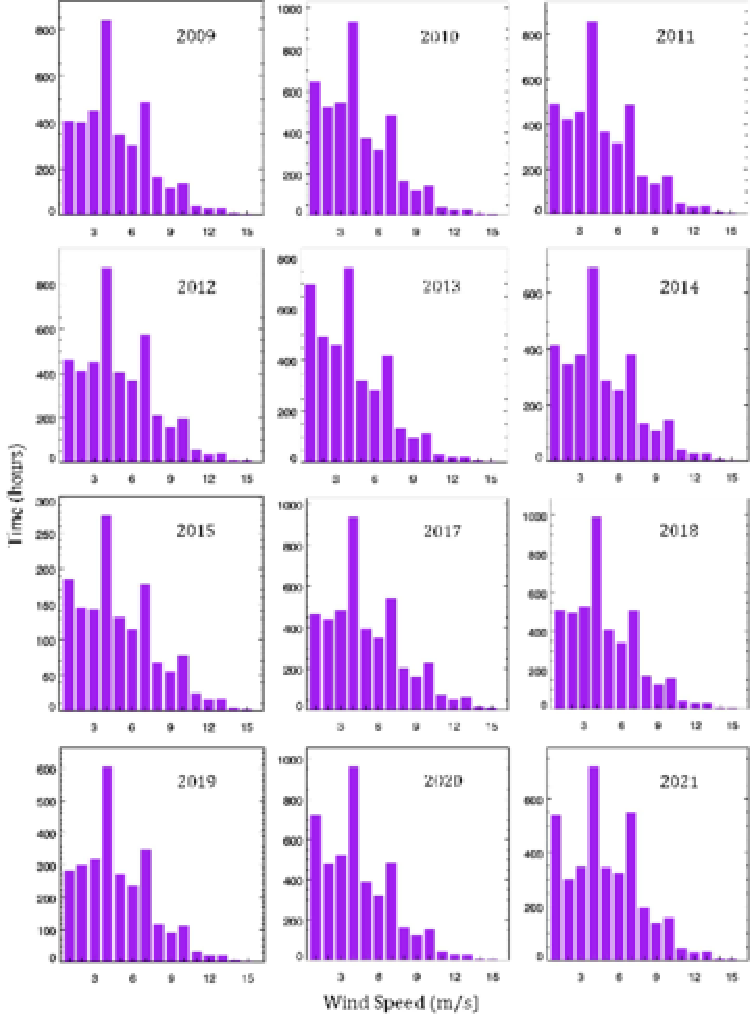}
  \caption{Histogram showing the yearly variation of wind speed from 2009 to 2021.}
  \label{fig:7}
\end{figure}

\begin{figure}[h!]
  \includegraphics[width=1\linewidth]{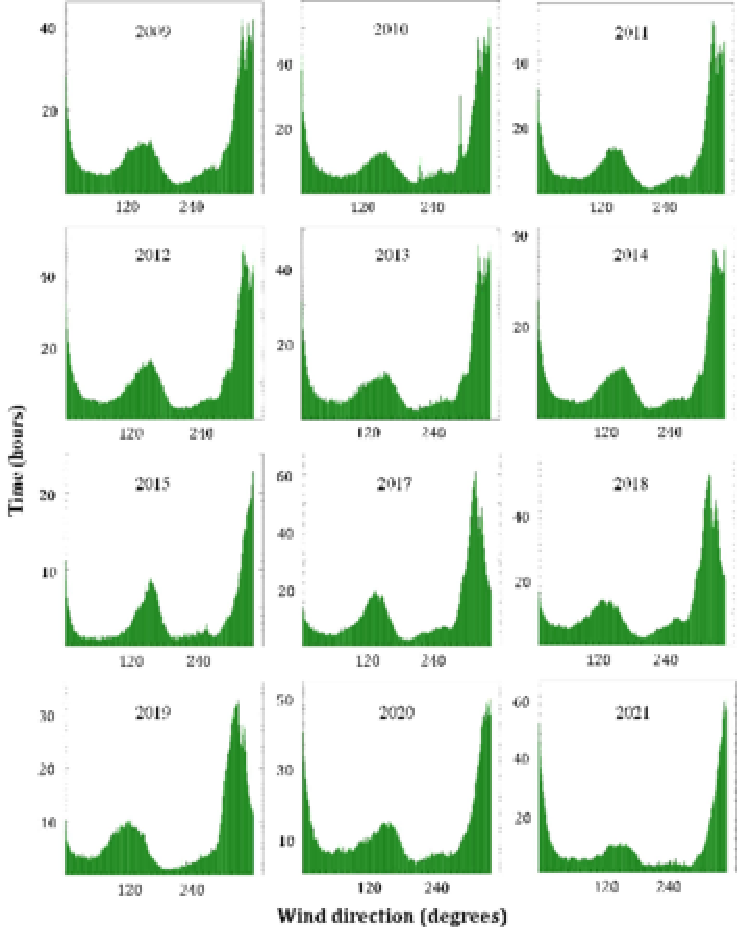}
  \caption{Distribution of wind direction during the day for the whole period of observation from the year 2009 to 2021.}
  \label{fig:8}
\end{figure}

\subsection{Wind Speed and Directions through Histograms}
It is clear from the wind-rose plot that the wind speed is usually less than 5~m~s$^{-1}$. To see this, we have computed the number of hours of wind blew in each bin with a step size of 1~m~s$^{-1}$. Figure~\ref{fig:7} shows the histogram of the same plotted for each year. The histogram shows that the first peak is at 4~m~s$^{-1}$ for all the years. The second peak is at 7~m~s$^{-1}$. The hills in the West and North-West direction may be slowing down the wind speed. When it reaches the lake it's speed becomes almost laminar and aligned along the lake direction with speeds between 4 and 7~ms~$^{-1}$. This generally occur from morning till noon. During this period good seeing is observed \citep{2019SoPh..294....5R, 2021SoPh..296...65U}. The tail of the wind speed extended beyond 15~m~s$^{-1}$. But their numbers are minimal. In all the years, about 65\% of the time, the wind speed falls between 1 and 5~m~s$^{-1}$. About 85\% of the time, the wind speed was less than 7~m~s$^{-1}$. 

Figure~\ref{fig:8} shows the distribution of wind direction in each year. Each bin along the X-axis is 1$^{\circ}$, and the Y-axis is in terms of number of hours each bin falls in the category. 60\% of the time, the wind is in the direction of 320$^{\circ}$, and about 25\% of the time it is in the direction of 160$^{\circ}$. This indicates the wind blows in the direction North-West most of the time and sometimes in the South-East direction.

\subsection{Temporal Variations of Wind Speed}
\begin{figure}[h!]
  \includegraphics[width=1\linewidth]{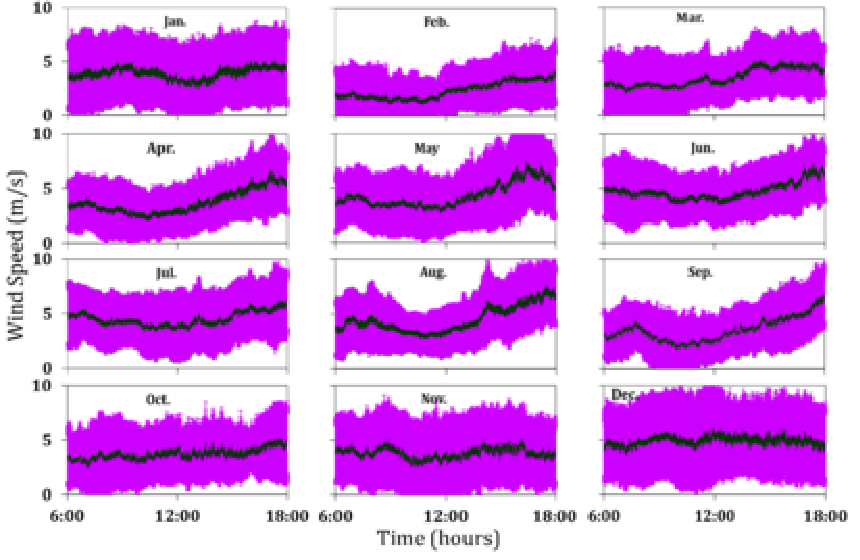}
  \caption{The variations of wind speed from morning to evening in each month of the year-2020. The black curve represents the mean value and the violet is the root mean square variations in the wind speed.}
  \label{fig:9}
\end{figure}

Figure~\ref{fig:9} shows the diurnal variation of wind speed in different months for the year 2020. This plot will show how the wind speed varies from morning to evening in each season at the site. The black curve represents the mean value, and the violet one is the RMS value. The wind speed is less than 5~m~s$^{-1}$ in the morning, and it becomes more than 5~m~s$^{-1}$ in the evening. In August, sometimes there is a sudden jump in the wind speed in the afternoon because of large wind speeds. The spread in the mean value is large over the winter months and a little less in the summer months. However, the wind speed is below 5~m~s$^{-1}$ most of the time. A similar trend in each month is found in all the years.

\begin{figure}[h!]
  \includegraphics[width=1\linewidth]{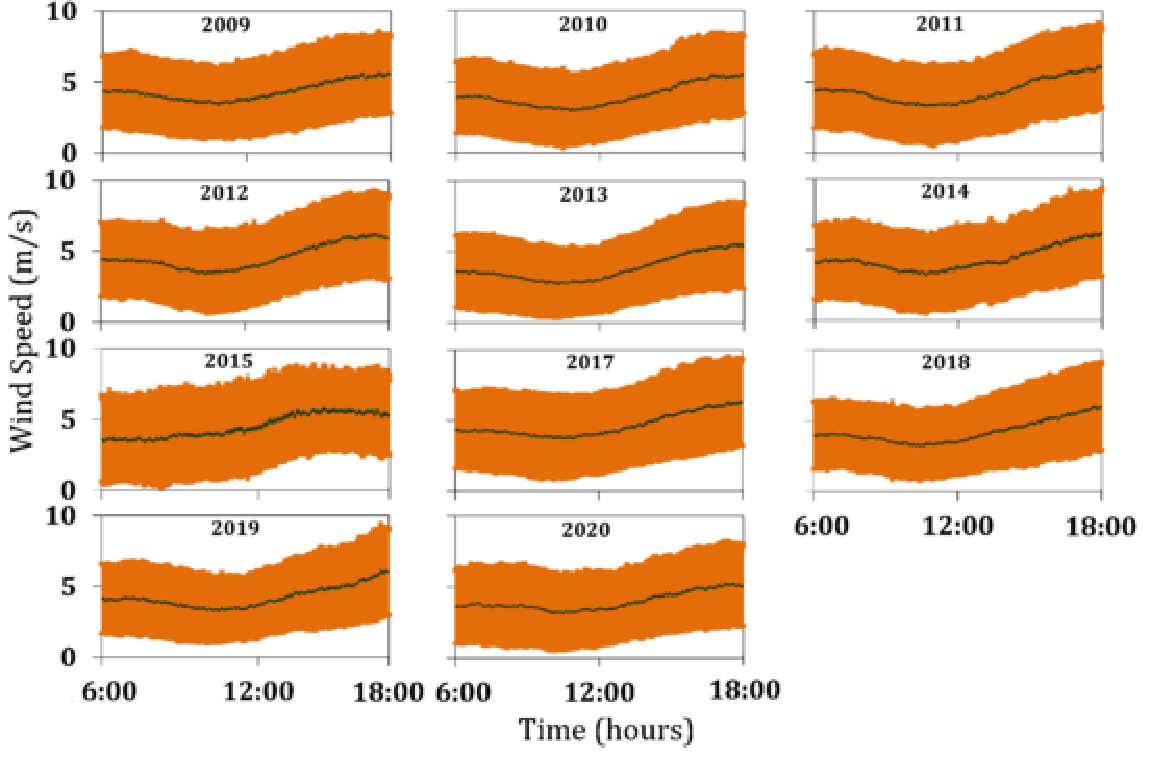}
  \caption{The variations of wind speed from monrning to evening averaged over the same time for the whole year is shown for each year. The black curve represents the mean value and the orange color represnts the RMS value.}
  \label{fig:10}
\end{figure}

Figure~\ref{fig:10} shows the {diurnal} variations of the wind speed from morning to evening on each year obtained by averaging the data from morning to evening of the same time on each day of the year. The plots are shown from 2009 to 2020. The black curve represents the mean value, and the orange color represents the spread over the mean value. In all these plots, the trend is the same. The wind speed during the morning hours is less than 5~m~s$^{-1}$, decreasing between 10 and 12 hours. After that, the wind speed increases. Except for 2015, the trend in the data is the same. In 2015 a few months data only available and hence the difference.

\subsection{Monthly/Seasonal Variations of wind speed}

\begin{figure}[h!]
  \includegraphics[width=1.0\linewidth]{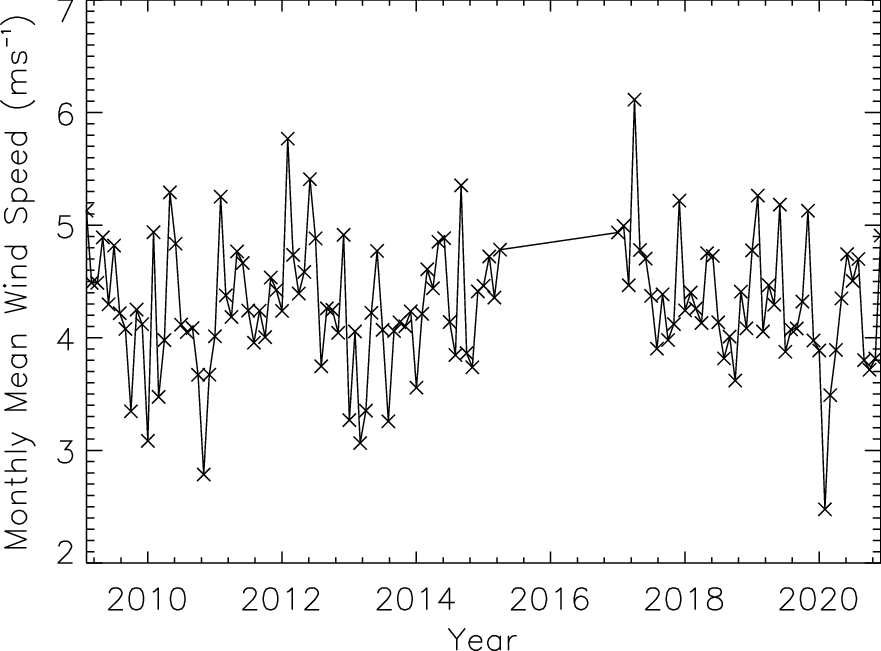}
  \caption{The seasonal variations of the mean wind speed from 2009 till 2021. There are 12 points for each year.}
  \label{fig:11}
\end{figure}

Due to seasonal variation of other meteorological parameters, wind speed also varied seasonally. The change in wind speed is caused by various factors over the season. We have plotted the monthly averaged wind speed from 2009 till 2021 in Figure~\ref{fig:11}. There are twelve points for each year. There is a data gap between 2015 and 2016 because of the unavailability of the AWS instrument at Merak. The average wind speed varies from 3 to 5 m~s$^{-1}$ over the season. In the winter months, the wind speed is a little larger than in the other months. However, there are not much variations in the wind speed in any of the seasons.

\begin{table*}[!h]
\centering
\caption{The number of hours of wind speed (ms$^{-1}$) falling in the range of speeds in each month of the year 2020 are tabulated.}
\begin{tabular}{|c|cccccccc|}
\hline
\multicolumn{9} {|c|}{Wind Speed (ms$^{-1}$)  in  number of  hours}\\
\hline

Month     & 1-3  & 3-5   & 5-7  & 7-9  & 9-11 & 11-13 & 13-15 & $>$15 \\\hline
Jan   & 74.3 & 123.9 & 61.6 & 24   & 4.4  & 0.3   & 0.02  & 0.0003   \\
Feb   & 97.7 & 98.3  & 38.1 & 3.8  & 0.12 & 0.004 & 0.0   & 0.0    \\
Mar   & 95.6 & 140.4 & 49.4 & 13.3 & 1.05 & 0.007 & 0.0003& 0.0   \\
Apr   & 98.8 & 144.1 & 77.9 & 12.3 & 1.2  & 0.016 & 0.0   & 0.0   \\
May   & 82.9 & 137.5 & 67.7 & 24.8 & 4.4  & 0.5   & 0.06  & 0.003 \\
Jun   & 54.8 & 154.2 &100.3 & 14.7 & 0.8  & 0.05  & 0.0   & 0.0   \\
Jul   & 71.4 & 175.7 & 80.9 & 18.4 & 4.1  & 0.3   & 0.06  & 0.0003 \\
Aug   & 75.1 & 181.0 &106.9 & 22.7 & 3.6  & 0.7   & 0.3   & 0.04  \\
Sep   & 87.8 & 165.4 & 57.3 & 17.9 & 0.75 & 0.03  & 0.003 & 0.0  \\
Oct   & 94.9 & 149.8 & 54.9 & 14.6 & 0.79 & 0.05  & 0.002 & 0.0  \\
Nov   &101.2 & 105.5 & 54.6 & 29.5 & 4.1  & 0.12  & 0.004 & 0.0  \\
Dec   & 85.1 & 138.1 & 76.6 & 48.8 & 12.9 & 0.97  & 0.03  & 0.001 \\\hline
Annual& 999.2&1670.6 &769.5 &225.9 & 36.4 & 2.4   & 0.2   & 0.008 \\
\hline
\end{tabular}
\label{tab:1}
\end{table*}

\begin{figure}[hbtp!]
  \includegraphics[width=1\linewidth]{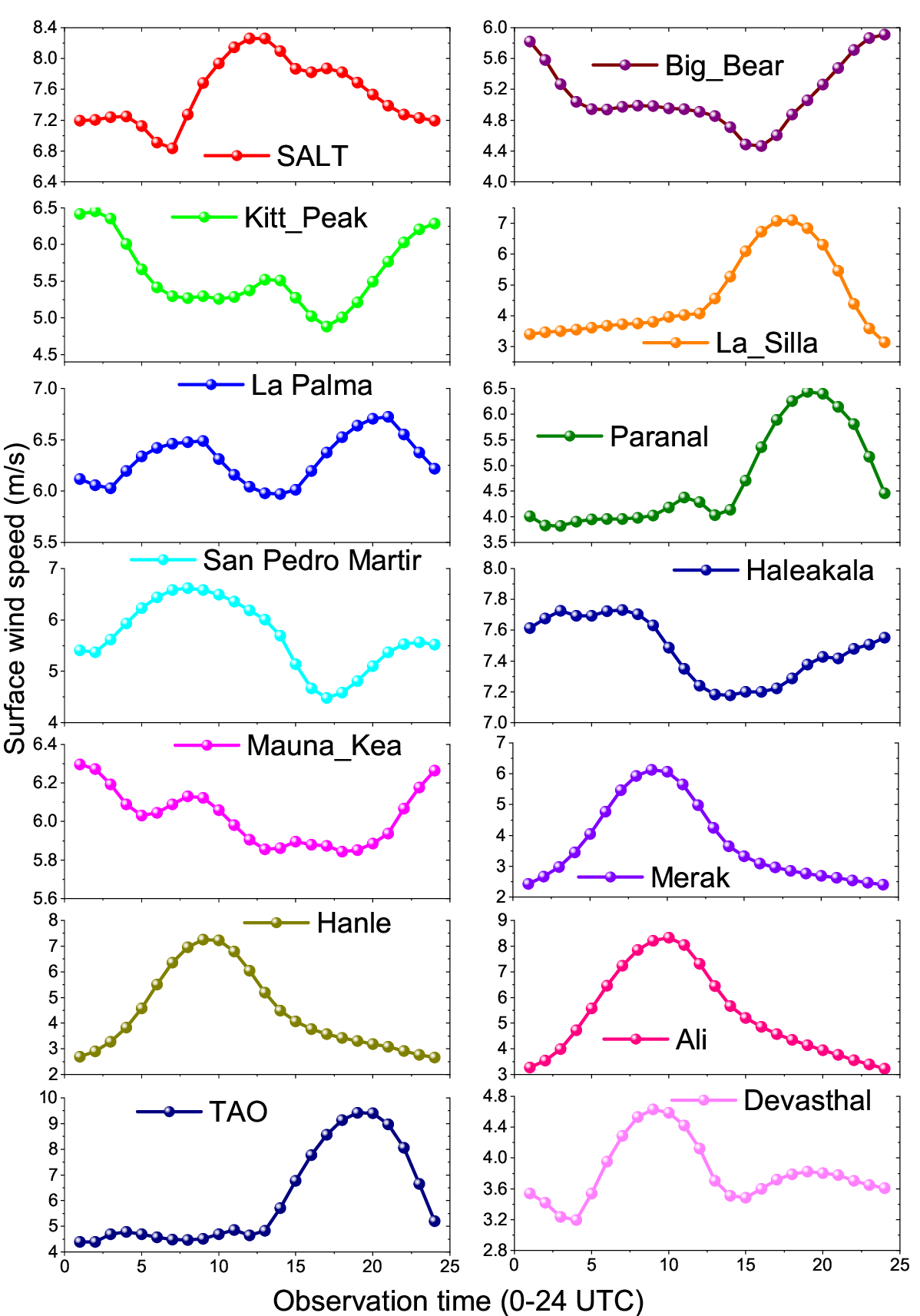}
 \caption{From top to bottom (left to right) shows the diurnal variation of surface wind speed at 14 astronomical sites listed in Table~\ref{tab:2} using hourly (0 to 24 hr UTC) data from MERRA-2 (0.50$^{\circ}$ $\times$ 0.625$^{\circ}$) during January 2020--December 2020.}
  \label{fig:12}
\end{figure}

Table~\ref{tab:1} shows the number of hours of wind speed falls in the range of 1-3, 3-5, 5-7, 7-9, 9-11, 11-13, 13-15, and larger than 15~ms$^{-1}$ in each month for the year 2020. Clearly, many hours fall in the wind speed range of 3-5~ms$^{-1}$. The second largest number of hours falls in the wind speed range of 1-3~ms$^{-1}$ and 5-7~ms$^{-1}$ depending on the month. For example, from January to May, large number of hours falls in 1-3~ms$^{-1}$, but from June to August, it is 5-7~ms$^{-1}$. There is a small number of hours when the wind speed is larger than 9~ms$^{-1}$. Similarly, annually, a large number of hours are falling in the wind speed range 3-5~ms$^{-1}$ as seen in Table~\ref{tab:1}. From the Table~\ref{tab:1} it is clear that about 95\% of the time wind speed is less than 7~m/s from February to June and it is about 90\% of from July to January. In all the months about 99.9\% of the time the wind speed is less than 15~m/s.

\section{Comparative studies among astronomical sites}

As a part of comparative study, 14 astronomical sites across the globe are examined using a homogeneous reanalysis data from MERRA-2. Figure~\ref{fig:12}, illustrates the diurnal variation of surface wind speed at 14 astronomical sites listed in Table~\ref{tab:2}. The data, sourced from MERRA-2 reanalysis (spatial resolution: 0.50$^{\circ}$ $\times$ 0.625$^{\circ}$) for January  to December 2020, is based on hourly wind speed measurements (m/s) from 0 to 24 UTC. A notable trend emerges: high-altitude sites exhibit the most significant wind speed variations between day and night. Ali and TAO, for instance, experience a 5~m/s difference, followed by Hanle (4.6 m/s), La Silla (4~m/s), and Merak (3.7~m/s). In contrast, low-altitude sites like Mauna Kea (0.5~m/s), Haleakala (0.6~m/s), and La Palma (0.8~m/s) show minimal diurnal variation.

Regarding average wind speed, TAO tops the list at 9.4 m/s, followed by Ali (8.3 m/s), SALT (8.26 m/s), Haleakala (7.7 m/s), Hanle (7.3 m/s), La Silla (7.1 m/s), Merak (6.1 m/s), and Devasthal (4.6 m/s). Figure~\ref{fig:12} presents the diurnal surface wind speed profile at these sites, derived from ground-based AWS data between 6:00 and 18:00 local time. Despite the coarser spatial resolution of the reanalysis data, a strong correlation is evident between the two datasets. This comparative analysis of diurnal surface wind speed at 14 astronomical sites, utilizing homogeneous reanalysis data, provides valuable insights into site-specific atmospheric conditions.

\begin{figure}[hbtp!]
\includegraphics[width=1.0\linewidth]{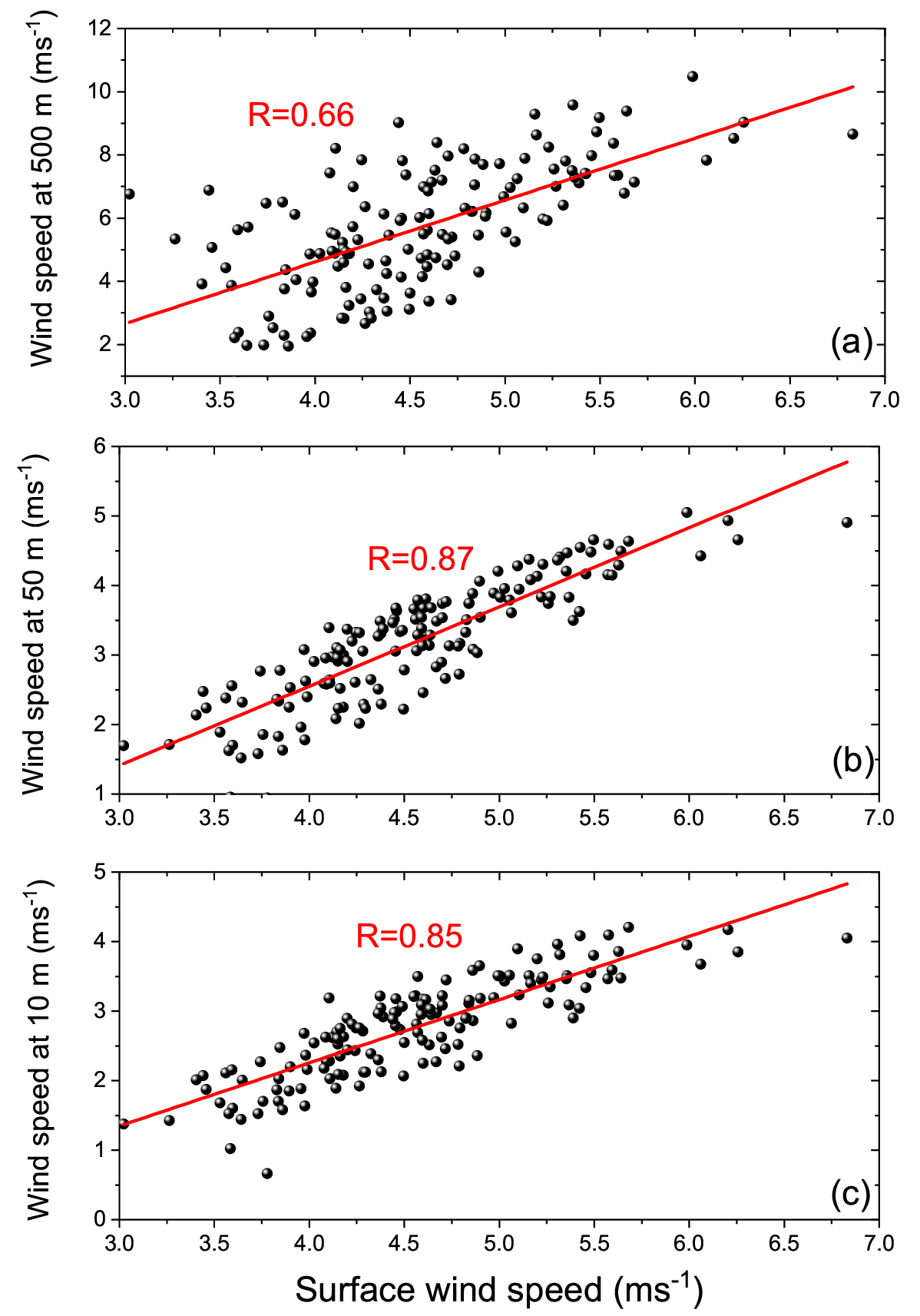}
\caption{Scatter plot of surface wind speed (ms$^{-1}$) with upper wind speed at 10~m, 50~m and 500~m above the surface. The correlation coefficient (R) between the two parameters are also shown in the Figure. The data shown in the Figure is monthly data at a spatial resolution of 0.5$^{\circ}$ by 0.625$^{\circ}$ during January 2009 to December 2020.}
\label{fig:13}
\end{figure}

\subsection{Wind gust and upper air wind profile}

A good astronomical site is decided from several astro-climatological parameters, mainly from good sky transparency (low aerosol content), high clear sky (low cloud coverage), and low turbulence or low wind gusts. A flat, uniform plain or gently rolling hills extending far upwind plain from the lake incursion site may bring a laminar wind flow at the proposed site. In earlier studies, \cite{2017ExA....43..145N} has reported wind gusts from several astronomical sites, including the present study location. According to their studies, the estimated wind gust at Big Bear, BBSO (34.25$^{\circ}$ N, 116.92$^{\circ}$ W, 2060 m amsl) is reported as the lowest among the stations, while Merak (33.80$^{\circ}$ N, 78.62$^{\circ}$ E, 4310 m amsl) and Kitt Peak, NSO (31.95$^{\circ}$ N, 111.58$^{\circ}$ W, 2098 m amsl) are found to be moderate with $\sim$11 ms$^{-1}$, and the highest wind gusts were experienced at Hanle (32.78$^{\circ}$ N, 78.95$^{\circ}$ E, 4500 m amsl) with median wind gusts at $\sim$15 ms$^{-1}$. Further, wind gust at Haleakala, HAO (20.71$^{\circ}$N, 156.26$^{\circ}$W, 3052 m amsl) lies $\sim$14 ms$^{-1}$, which is lower than Merak as reported therein. The strength of the wind gust varies with altitude (spatial) and temporally. For example, wind gusts during the daytime are higher than at night time, which is attributed to thermal convection from the warmer surface during the daytime.

\begin{center}
\begin{table}[tp]
\caption{Location of selected astronomical sites for upper wind at 200 mbar by using MERRA-2 monthly reanalysis data during 2008-2021.}
\begin{tabular}{llllll}\\\hline\hline
Sites            &Long.     &Lat.   & Alt.  &   Mean  & Country  \\
SALT             &  20.81   &-32.38 &  1798 &   31.39 & S. Africa\\
Big\_Bear        &-116.92   & 34.25 &  2060 &   20.85 & USA      \\
Kitt\_Peak       &-111.58   & 31.95 &  2098 &   22.78 & USA      \\
La Silla         & -70.73   &-29.25 &  2400 &   34.76 & Chile    \\
La Palma         & -17.88   & 28.77 &  2400 &   18.76 & Spain    \\
Devasthal        &  79.68   & 29.36 &  2450 &   30.22 & India    \\
Paranal          & -70.40   &-24.63 &  2635 &   29.12 & Chile    \\
San Pedro Martir &-115.45   & 31.03 &  2800 &   21.41 & Mexico   \\
Haleakala        &-156.26   & 20.71 &  3052 &   19.68 & USA      \\
Mauna Kea        &-155.47   & 19.83 &  4100 &   19.69 & USA      \\
Merak            &  78.62   & 33.8  &  4310 &   30.99 & India    \\
Hanle            &  78.96   & 32.78 &  4500 &   31.32 & India    \\
Ali              &  80.00   & 32.63 &  5100 &   31.32 & China    \\
TAO              & -67.74   &-22.98 &  5640 &   25.50 & Chile    \\
\hline
\end{tabular}
\label{tab:2}
\end{table}
\end{center}

Generally, wind speed decreases towards the surface due to lesser energy associated with frictional wind flow from the rough surface. The amplitude of atmospheric turbulence mainly decides the quality of ground-based astronomical observatories. The angular resolution of an optical telescope is determined by its diameter, which is critically affected by high atmospheric turbulence. In this aspect, high-altitude wind speed at 200-mbar pressure level plays a vital role in examining the amplitude of the atmospheric turbulence as reported by \cite{2005MNRAS.356..849G, 2002ESOC...58..321S}. In earlier studies, \cite{2017ExA....43..145N} shows a good relation (R=0.79) with night time seeing and upper air wind at 400~mbar pressure level over IAO-Hanle. However, there are weak or negative correlations at specific heights in the atmosphere, as reported therein, which may be associated with the dynamics of the atmosphere in different atmospheric layers.

Further, \cite{2002ESOC...58..321S} found a good correlation between the average speed of the turbulence and wind speed at 200~mbar pressure level. In the current work, we explored how the low-altitude atmospheric winds correlate well with the upper-air wind speed. Figure~\ref{fig:13} shows the correlation of surface wind speed from MERRA-2 with the upper air wind speeds at 10 m, 50 m, and 500 m above the surface. The wind speed data shown in the Figure is the monthly data at a spatial resolution of 0.5$^{\circ}$ by 0.625$^{\circ}$ from January 2009 to December 2020. The surface wind speed correlates well with the upper air wind speed with the correlation coefficients of 0.85, 0.87, and 0.66 at 10 m, 50 m, and 500 m above the surface, respectively. Surface wind at the lower surface shows a better correlation, apparently disturbing the linear relation at the higher level. 

The current work also examined high-altitude wind at 200~mbar pressure level at several astronomical sites across the globe since the wind speed at 200~mbar pressure level is one of the essential parameters to correlate with the turbulence in the atmosphere (\citep{2005MNRAS.356..849G}). Table~\ref{tab:2} shows wind speed profile at 200~mbar pressure level from monthly MERRA-2 data during 2008-2020 at fourteen important astronomical observatories. The spatial resolution of the data is taken at 0.5$^{\circ}$ by 0.625$^{\circ}$ at 73 vertical pressure levels. It is found that Merak's wind speed at 200~mbar pressure level is very close to 31~ms$^{-1}$ and for Hanle, Ali, and SALT, are shown in the Table. The minimum and maximum wind speeds are observed at La Palma in Spain and La Silla in Chile with 18.67~ms$^{-1}$ and 34.76~ms$^{-1}$, respectively. According to the statistics of high altitude wind at 200~mbar pressure level, La Palma and La Silla have the lowest and highest wind speed, which appears to be the most suitable and least suitable sites, respectively, among the selected fourteen astronomical sites.

\begin{figure}[hbtp!]
\includegraphics[width=1.0\linewidth]{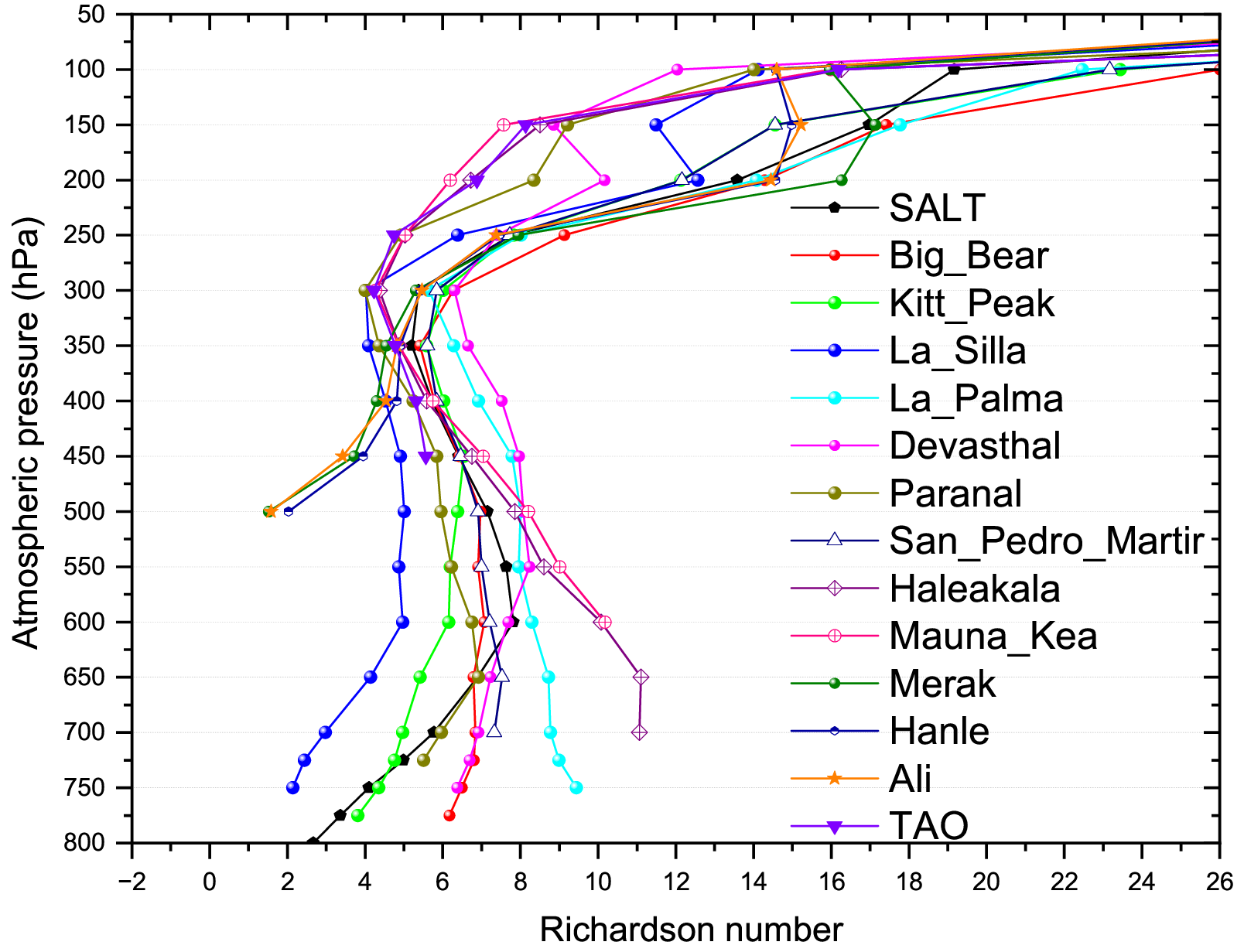}
\caption{Vertical profile of  Richardson number at 14 astronomical sites from monthly MERRA-2 (0.50 $\times$ 0.6250) data during January 2020--December 2020. }
\label{fig:14}
\end{figure}

Atmospheric turbulence significantly impacts the seeing conditions at astronomical sites. Assessing turbulence data from diverse sites over specific time frames is challenging due to data heterogeneity and limited public availability. To address this, we utilize a key turbulence parameter known as Richardson number in the current work from the MERRA-2 global reanalysis data during January 2020 to December 2020. It is a dimensionless quantity and calculated based on upper-air thermodynamic profiles, influenced by turbulent heat fluxes (Town and Walden, 2009) and optical turbulence (Yang et al., 2022). Lower values indicate higher turbulence, while higher values signify a more stable atmosphere. Typically, values between 10 and 0.1 are considered, with values below unity suggesting significant turbulence. 

High-altitude mountain regions are prone to atmospheric turbulence. Figure~\ref{fig:14} illustrates the Richardson number profiles at 14 astronomical sites listed in Table~\ref{tab:2}. Due to the limited vertical resolution of reanalysis data (varying from 25 to 50 hPa), there are inherent technical limitations. The study found that Hanle, Merak and Ali, located in the high-altitude north-western Himalaya, have the most similar Richardson number (1.5 to 2.0) at 500 hPa. Devasthal, the another astronomical site located in the foot-hills of the Himalaya, shows Richardson number 6.4 at 750 hPa. Further, La Silla, the another astronomical site, located in the outskirt of the Atacama desert in Chile, has Richardson number 2.14 at 750 hPa, which is also close to the three Himalayan sites (Hanle,Merak and Ali). These comparative studies of Richardson number show the  turbulence parameter at the current study location.

\section{Summary and Discussions}
\label{sec:5}
We have presented the results of the wind pattern seen in the incursion site of Pangong Lake. We have analyzed the weather data at the site from June 2008 to 2020, with a gap of 1.5~years between 2015 and 2016. The monthly averaged and yearly averaged data shows the wind blew in the direction of the northwest most of the time. The wind speed is less than 5~ms$^{-1}$ most of the time, and there is another peak at 7~ms$^{-1}$ as well. In all the seasons, the wind speed is less than 5~ms$^{-1}$ in the morning hours, and it increases by a few ms$^{-1}$ in the evening. The trend of the wind pattern is the same in all the years, with some seasonal effects. 

The Pangong Lake is running in the N-S direction. The wind direction is also the same, which is along the lake most of the time of the year. The wind speed is mild, below 5~ms$^{-1}$ in that direction, which helps in improving the seeing at the site (NLST site survey report). In the furture, we will compare the results of solar seeing at the site with the wind speed and direction.

 Surface wind speed correlates well with the high-altitude wind speeds at 10 m, 50 m, and 500 m with correlation coefficients of 0.85, 0.87, and 0.66, respectively. Further, according to the statistics of high altitude wind at 200~mbar pressure level, La Palma is found to be the most suitable site with wind speed 18.76~m$^{-1}$ and La Silla as the bottom line with wind speed 34.76~ms$^{-1}$ in the selected astronomical sites. The estimated wind speed at Merak is 30.99~ms$^{-1}$.
 
The speed between 4 and 7 m/s is considered as moderate wind speed which helps to stabilize the atmosphere. A complete lack of wind can lead to stagnant air pockets, where temperature differences can build up and create significant turbulence.  The speed between the aforesaid values are optimum for solar observations because a gentle breeze helps to average out the temperature differences within the turbulent cells (caused by temperature variations), reducing the severity of the refractive index variations that cause the distortion. While moderate winds can help break up turbulent cells, strong winds can actually generate new turbulence, particularly near the ground. This can lead to increased atmospheric distortion and reduce the image clarity.

In the location, a large number of hours the wind speed is between 3 and 5~m/s. This range of wind speed is found generally in the morning hours and in almost all the season. The next one is 1-3~m/s.  This wind speed range is observed mostly in the winter season, from November to April. Instead in the summer season 5-7 m/s wind speed is observed. December and January months are season where certain number of hours the wind speed is larger than 9 m/s. This number is small in the summer, but it prevails in May, July and August. From Table~\ref{tab:1}, we can conclude that summer season is best where large number of hours the wind speed is between 3 and 5~m/s with mild gusts. In 2015, data was only available for a few winter months. While the annual wind direction remained consistent, the morning-to-evening wind speed pattern differed from other years.
  
Our analysis reveals that Hanle, Merak, and Ali, situated in the high-altitude northwestern Himalayas, exhibit similar Richardson numbers (1.5 to 2.0) at 500 hPa. Devasthal, located in the Himalayan foothills, shows a Richardson number of 6.4 at 750 hPa. Additionally, La Silla, situated in the Atacama Desert in Chile, has a Richardson number of 2.14 at 750 hPa, closely resembling the values observed at the three Himalayan sites. This comparative study provides valuable insights into turbulence parameters at these astronomical sites.

The land where the wind measurements are made is protruding into the water body. The wind is channelled by the two adjoining parallel mountain ranges and it flows over the water body for most part of the day. The wind speed is steady between 3 and 7~m/s with mild gusts. This was identified as major advantage of the site. It was seen that during this period the stable H-alpha images \citep{2018JApA...39...60R, 2019SoPh..294....5R, 2021SoPh..296...65U} are obtained suggesting that this range of wind speed improves the seeing. The North-south orientation of the valley and the lake acts as a channel for winds. The N-W winds will travel for several kilometers over the open waters of Pangong Lake before they reach the incursion site. Since the winds do not encounter any obstructions, the air flow is mostly laminar and the lake acts as a heat reservoir or heat sink. 

About 90\% of the time the wind speed is less that 7 m/s and 99\% of the time the wind speed is less than 15~m/s. This shows that the wind speed is mild most of the time at the Merak site. This type of wind speed is seen in many observatory locations \citep{Sayed2017, 2000A&AS..147..271J, 2006PASP..118..172B, 2006SPIE.6267E..0LD}. To the end, the Merak site is good for solar observations with a wind speed is less than 7~m/s most of the time and the direction is along the lake makes it better seeing at the site. 

The study found that Hanle, Merak and Ali, located in the high-altitude north-western Himalaya, have the most similar atmospheric turbulence profile, assigned by Richardson number with 1.5 to 2.0 at 500 hPa. The estimated parameter is close to 2.14 at 750 hPa over La Silla, another astronomical site, located in the outskirt of the Atacama desert in Chile. It is also necessary to examine other meteorological parameters from Merak with the other sites located across the globe. Therefore, in future, we will analyze all other parameters at the site and compare them with other astronomical sites around the world.


\section*{Acknowledgements}
We thank the reviewer for fruitful comments on the orginal version of the manuscript.
The project was initiated by S.~S.~Hasan, former director of Indian Institute of Astrophysics. In the intial stages, the site survey instruments were set up by S.~P.~Bagare, K.~E.~Rangarajan, A.~K.~Saxena, and T.~P.~Prabhu. The help of P.~A~ Samson, Md. Ismail, R. Ismail Jabillullah, T. G. Adiya, Rajendra B. Singh, Sangita K Padhy, N. Vasantharaju, K. Anjali John, B. Prabhu Ramkumar, Tsewang Dorje, are greatly acknowledged. The help of guest observers K.~P.~Raju, K.~B.~Ramesh, A.~Satyanarayana and Wahabuddin was useful. The help of Tsering Chonzom, Rigzin Namgyal, Tundup Thinless, Tashi Namgyal and Thinless Namgail is greatfully acknowledged. The support by various department of IIA, IAO, and ARIES are greatfully acknowledged.

\section{Author's Contribution}
Tundup Stanzin, Namgyal Dorje and Angchuk Dorje have collected the data for several years.
Deepangkar Sarkar and Shantikumar Singh analyzed the data.
Ravindra B and Shantikumar Singh wrote the manuscript text.
All other authors reviewed and commented on the manuscript.

\section{Funding}
Indian Institute of Astrophysics provided necessary logistics and support to carryout the wind
measurements at Merak.

\section{Data availability of Materials}
Data will be shared on request.

\section{Conflict of Interest}
Authors do not have any competing interests as defined by Springer, or other interests that might be percieved to influence the results and/or discussion reported in this paper.

\bibliography{ms_biblio}

\end{document}